\documentclass{ptapap}

\usepackage{xcolor}

\usepackage{soul}

\definecolor{dr}{rgb}{0.8,0,0}

\definecolor{db}{rgb}{0.1,0.1,0.8}

\definecolor{dg}{rgb}{0.,0.6,0.1}

\definecolor{dm}{rgb}{0.7,0.,0.7}

\definecolor{dgr}{rgb}{0.4,0.4,0.4}

\newcommand{\be}{\begin{equation}}
\newcommand{\ee}{\end{equation}}
\newcommand{\bea}{\begin{eqnarray}}
\newcommand{\eea}{\end{eqnarray}}
\newcommand{\bse}{\begin{subequations}}
\newcommand{\ese}{\end{subequations}}

\newcommand{\F}{{\mathcal F}}

\author{Magdalena Sieniawska}[CAMK]
\author{Michał Bejger}[CAMK]
\author{Paweł Ciecieląg}[CAMK]
\author{Andrzej Królak}[IMPAN]

\affil[CAMK]{Nicolaus Copernicus Astronomical Center, Polish Academy of Sciences, Bartycka 18, 00--716 Warszawa, Poland}
\affil[IMPAN]{Institute of Mathematics, Polish Academy of Sciences, Śniadeckich 8, 00--656 Warszawa, Poland}

\title{Followup procedure in time-domain $\F$-statistic searches for continuous gravitational waves} 
\headtitle{Continuous GW searches followup procedure}

\begin{document}

\maketitle

\begin{abstract}
Potentially interesting gravitational-wave candidates (outliers) from the blind
all-sky searches have to be confirmed or rejected by studying their origin and
precisely estimating their parameters. 
We present the design and first results
for the followup procedure of the {\tt Polgraw} all-sky search pipeline: a coherent
search for almost-monochromatic gravitational-wave signals in several-day long
time segments using the $\F$-statistic method followed by the coincidences
between the candidate signals. Approximate parameters resulting in these two
initial steps are improved in the final followup step, in which the signals from detectors are studied separately, together with the network
combination of them, and the true parameters and signal-to-noise values are established.
\end{abstract}

\section{Introduction}
\label{sect:intro} 

The all-sky time-domain $\F$-statistic pipeline\footnote{Project
repository and documentation: {\tt https://github.com/mbejger/polgraw-allsky}} \citep{jks98}
aims at detecting almost monochromatic continuous gravitational-wave (GW) signals of
a frequency $f$ and its time derivative $\dot{f}$.  

An astrophysical source providing a periodically time-varying mass or current
quadrupole is e.g. a non-axisymmetric rotating, spinning-down neutron star. The deviation
from axisymmetry may be caused by elastic or magnetic stresses, instabilities,
accretion or non-equilibrium heating  \citep[see e.g.][for a review]{Lasky2015}. 

The {\tt Polgraw} all-sky search pipeline is composed of the following components:
(I) preparation of the initial narrow-band time-domain
series (typically $1$ Hz width and a few days long, with the sampling rate of the order of 1 s) 
and ephemerides needed for the  position of the detectors with respect to the 
source, and the optimal grid of templates \citep{PisarskiJ2015} in the
parameter space of frequency $f$, frequency derivative (spindown) $\dot{f}$,
right ascension $\alpha$ and declination $\delta$, (II) the {\tt search} code i.e.: a coherent search for candidate signals in narrow-band time-domain segments of length
$T\simeq$ few days (definition of the $\F$-statistic is given in \citealt{jks98}.

\begin{figure}
  \centering
  \begin{minipage}{0.48\textwidth}
    \includegraphics[width=\textwidth]{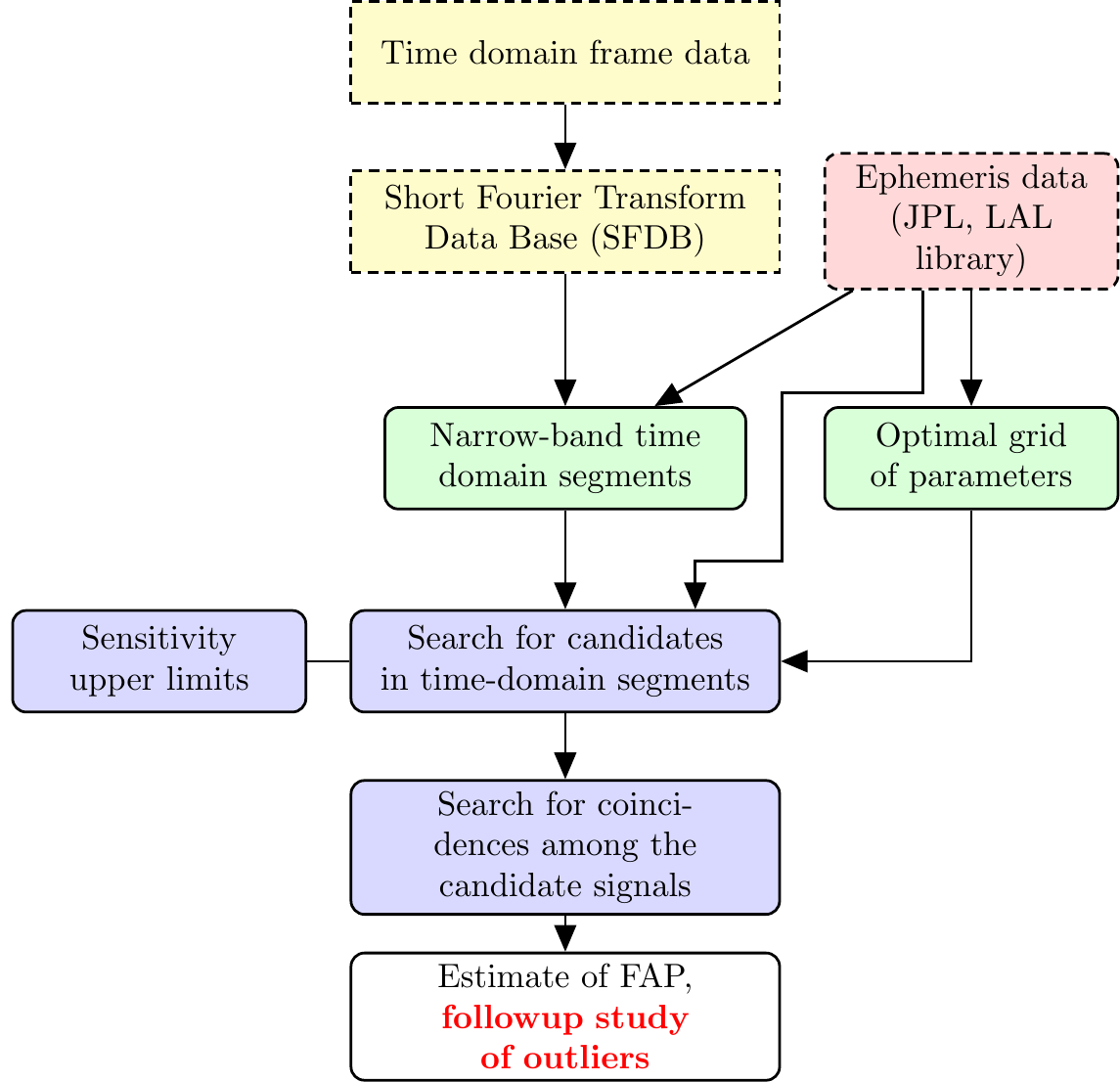}
    \caption{Block diagram of the time-domain $\F$-statistic pipeline.}
    \label{fig:pipeline}
  \end{minipage}
  \quad
  \begin{minipage}{0.48\textwidth}
 First implementation of the all-sky algorithm was presented in
\citet{AstoneBJPK2010}. Currently the data from a network of $N$ detectors
may be combined coherently, resulting in the sensitivity improvement
proportional to $\sqrt{N}$ \citep[see][]{Abbott2017a}, (III) {\tt
coincidences} code i.e.: incoherent search for coincidental candidate signal between the
time-domain segments  \citep[see][and the code documentation for
detailed description]{VSR1TDFstat}, (IV) calculation of the sensitivity upper limits by software
injections ({\tt sensitivity-scripts}), (V) estimation of the False Alarm
Probability of coincidences (FAP) and the {\tt followup} of promising
coincidences. Block diagram of the pipeline is presented in Fig.~\ref{fig:pipeline}.  
  \end{minipage}
\end{figure}

The pipeline has been used in the all-sky search for periodic GWs in the Virgo VSR1 data \citep{VSR1TDFstat}, LIGO O1 data \citep{Abbott2017a} and in the LIGO S6 Mock Data Challenge \citep{AllSkyMDC}. For a recent summary of search methods used 
by the LIGO/Virgo collaboration see e.g.: \citet{Bejger2017}. 

\section{The {\tt followup} procedure}
\label{sect:followup} 

\begin{figure}
  \centering
  \begin{minipage}{0.48\textwidth}
    \includegraphics[width=\textwidth]{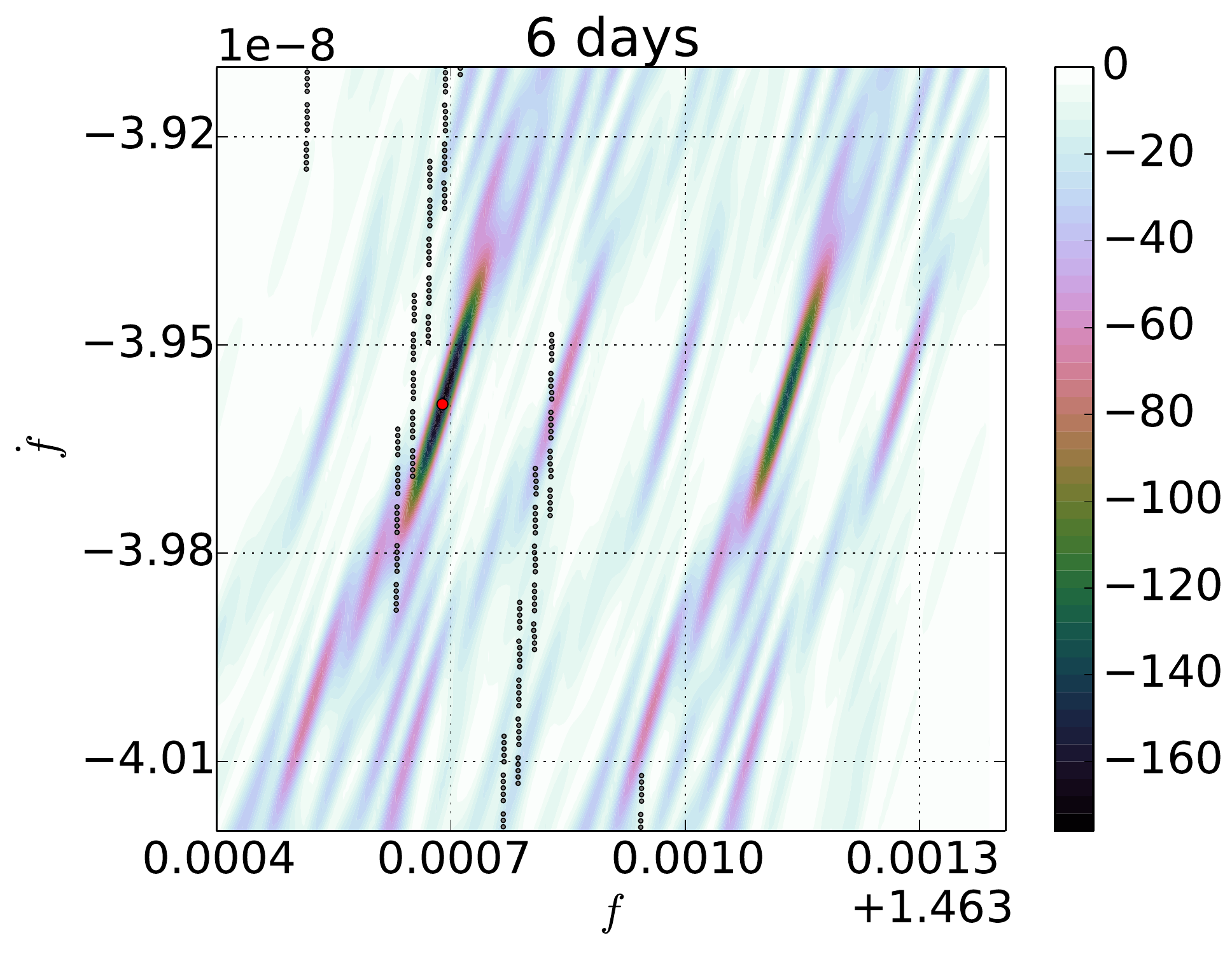}
    \caption{Shape of the $\F$-statistic in frequency-spindown plane, with an injected maximum (red point) and points of optimal grid (gray points).}
    \label{fig:konturandgrid}
  \end{minipage}
  \quad
  \begin{minipage}{0.48\textwidth}
    \includegraphics[width=\textwidth]{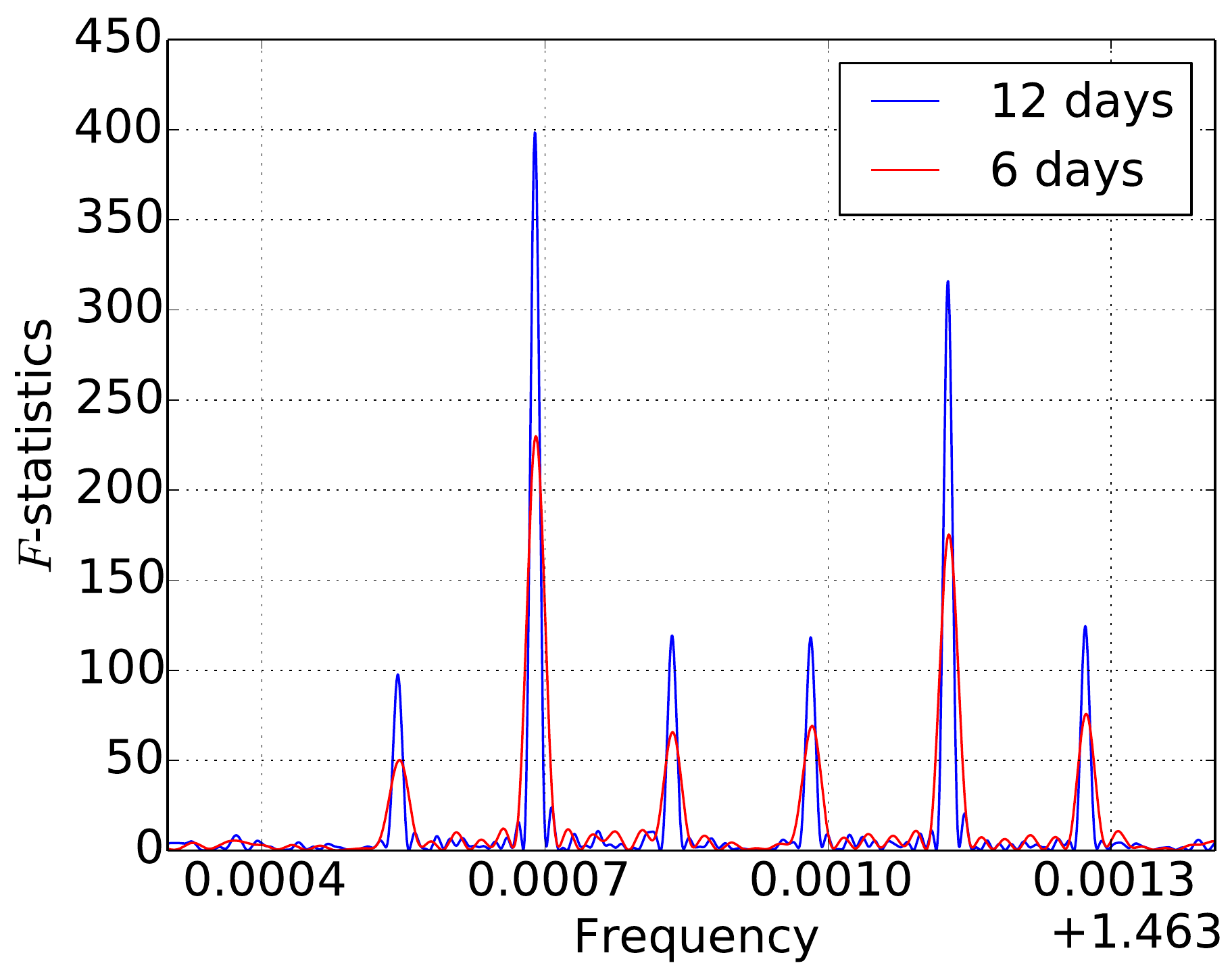}
    \caption{Shape of the $\F$-statistic as a function of the frequency of the 6- and 12-days data segments.}	
    \label{fig:fstatshape}
  \end{minipage}
\end{figure}

The {\tt followup} code implements the last stage of the time-domain $\F$-statistic
pipeline: final validation to confirm the astrophysical origin of the promising
candidates, establish their signal-to-noise ratio (SNR) and estimate their
parameters ($f$, $\dot{f}$, $\delta$, $\alpha$). Input data consists of time
series (divided into 6-days segments), detector ephemeris and optimal grid
(much denser than in the {\tt search} code). Main goal of the code is to find
maximum of $\F$-statistic in the optimal and fastest way. Here is the
description of the {\tt followup} procedure. In stage 1a a promising (low FAP) candidate
from {\tt coincidences} results is selected, and the closest point on a dense optimal
grid is found. Then at some nearby grid points (e.g., $\pm$ 2 points in all
directions $\to$ 625 points) the $\F$-statistic is evaluated (employing OpenMP
parallelization). Point with the highest $\F$-statistic (highest SNR) is the
initial point for stage 1b that consists of direct maximum search by either
Simplex Nelder--Mead method \citep{neldermead} or MADS  \citep[Mesh Adaptive Direct
Search;][]{mads} algorithm. Stage 1 is repeated for next data segment. Stage
2 consists of concatenating adjacent data segments and their ephemerides,
calculating weighted (by the SNR) mean of results and taking it as an initial
point; finally, running {\tt followup} for the concatenated data (directly with
Simplex or MADS). Fig.~\ref{fig:konturandgrid} contains an example of the non-trivial 
structure of the $\F$-statistic shape.  

\begin{figure}
\begin{center}
    \includegraphics[width=0.8\textwidth]{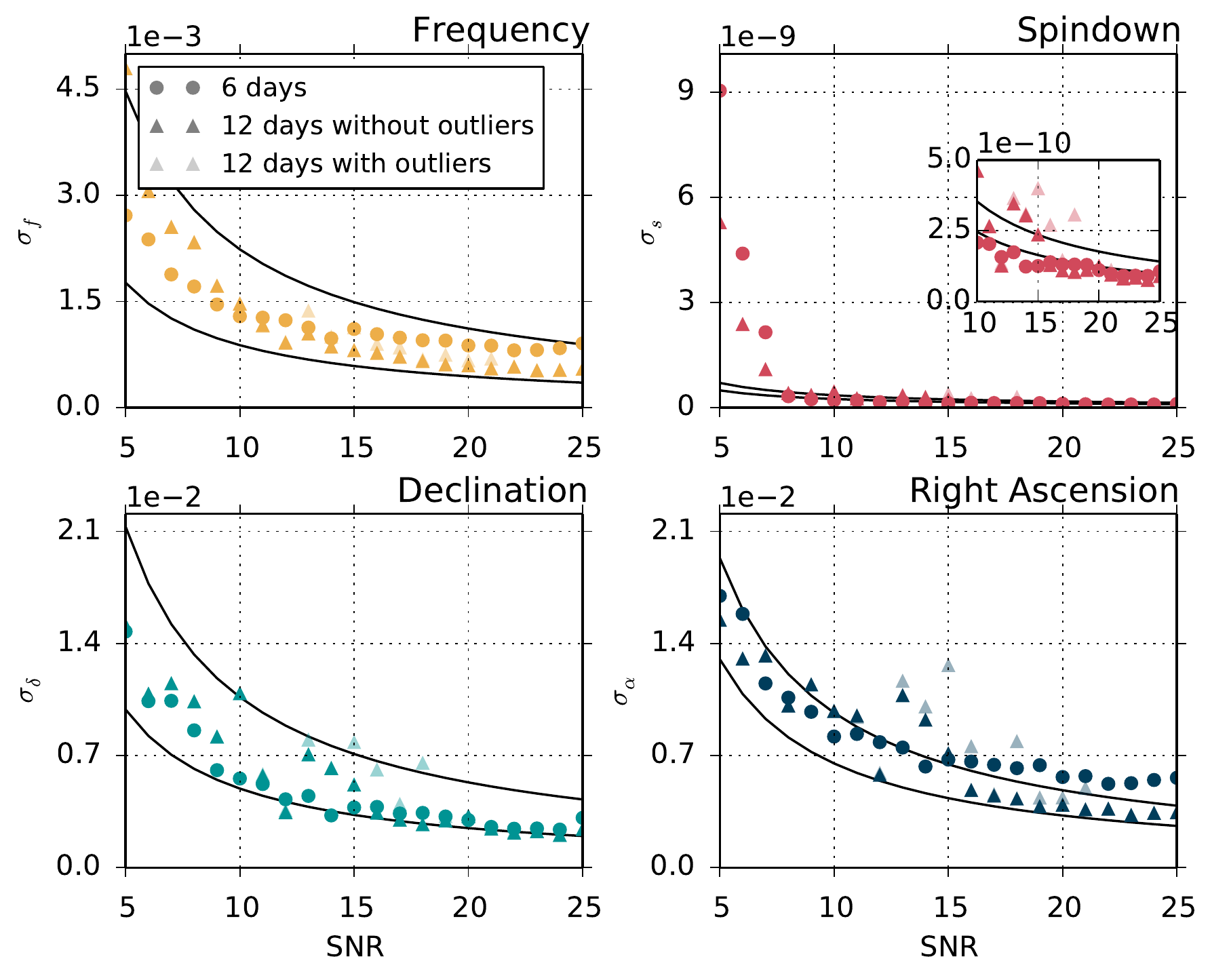}
    \caption{Standard deviations for 6- and 12-days segments and corresponding theoretical predictions (Cram{\'e}r-Rao bounds) as black lines: upper for 6 days, lower for 12 days. In a few cases local maxima instead of global ones were found by the {\tt followup} (they were identified and removed not to spoil the comparison).}
    \label{fig:std_6and12d}
\end{center}     
\end{figure} 

\section{Numerical challenges} 
\label{sect:numchall} 

Because the computational cost of the {\tt search} code scales with the segment
length as $\propto T^5\log T$ \citep{AstoneBJPK2010}, one needs to divide data
into shorter chunks, like 6-days segments. At the {\tt followup} stage, when
one can study the focus only on very narrow part of the parameter space, longer
data series are possible. The SNR value increases with the segment length as
$\propto \sqrt{T}$, so in principle it is easier to find the maximum of the
$\F$-statistic. Alas, with the increasing SNR, some of numerical problems
become more vivid: not only the global maximum, but also the local maxima increase. Moreover,
all maxima become narrower, therefore when a maximum is found, it is not straightforward
to establish whether it is a global or local one. In fact, for concatenated
frames, it's easier to find any maximum, but it is more difficult to find a
global one (example for concatenated data is presented in Fig.~\ref{fig:fstatshape}).
Results presented in Fig.~\ref{fig:std_6and12d} were obtained by adding a
software injection (an artificially generated signal of a given SNR) to a
sample of 250 realizations of the Gaussian noise, detecting it with the {\tt
followup} procedure and comparing the recovered parameters to the injected
signal parameters. 

\section{Conclusions}
\label{sect:conclusions} 

We conclude that the {\tt followup} procedure is able to estimate the parameters of
injected signals with a satisfactory accuracy, as demonstrated by the
comparison with the theoretical predictions. There is still some room for
improvement (e.g., with the MADS method). The final goal is to effectively
deploy the {\tt followup} procedure on hierarchically-concatenated data
consisting of the whole observational run ($\simeq$ 1 year of data). 

\acknowledgements{This work was partially supported by the Polish NCN grants no.
2014/14/M/ST9/00707 and 2016/22/E/ST9/00037.} 

\bibliographystyle{ptapap}
\bibliography{sieniawska}

\end{document}